# CLUSTERING WITH OBSTACLES IN SPATIAL DATABASES

Mohamed A. El-Zawawy [1]          Mohamed E. El-Sharkawi [2]

[1]Dept. of Mathematics
Faculty of Science
Cairo University
Cairo, Egypt
mzawawy@operamail.com

[2]Dept. of Information Systems
Faculty of Computers & Information
Cairo University
Cairo, Egypt
mel_sharkawi@hotmail.com

## ABSTRACT

*Clustering large spatial databases is an important problem, which tries to find the densely populated regions in a spatial area to be used in data mining, knowledge discovery, or efficient information retrieval. However most algorithms have ignored the fact that physical obstacles such as rivers, lakes, and highways exist in the real world and could thus affect the result of the clustering. In this paper, we propose CPO, an efficient clustering technique to solve the problem of clustering in the presence of obstacles. The proposed algorithm divides the spatial area into rectangular cells. Each cell is associated with statistical information used to label the cell as dense or non-dense. It also labels each cell as obstructed (i.e. intersects any obstacle) or non-obstructed. For each obstructed cell, the algorithm finds a number of non-obstructed sub-cells. Then it finds the dense regions of non-obstructed cells or sub-cells by a breadth-first search as the required clusters with a center to each region.*

## 1. INTRODUCTION

Spatial databases contain spatial-related information such databases include geographic (map) databases, VLSI chip design databases, and medical and satellite image databases. Spatial data mining is the discovery of interesting characteristics and patterns that may exist in large spatial databases. It can be used in many applications such as seismology (grouping earthquakes clustered along seismic faults) and geographic information systems (GIS). Clustering, in spatial data mining, is a useful technique for grouping a set of objects into classes or clusters such that objects within a cluster have high similarity among each other, but are dissimilar to objects in other clusters. Many effective clustering methods have been developed. Most of these algorithms, however, dose not allow users to specify real life constraints such as the existence of physical obstacles, like mountains and rivers. Existence of such obstacles could substantially affect the result of a clustering algorithm. For example, consider a telephone-company that wishes to locate a suitable number of telephone cabinets in the area shown in Figure 1-a to serve the customers who are represented by points in the figure. There are natural obstacles in the area and they should not be ignored. Ignoring these obstacles will result in clusters like those in Figure 1-b, which are obviously inappropriate. Cluster $cl_1$ is split by a river, and customers on one side of the river have to travel a long way to reach the telephone cabinet at the other side. Thus the ability to handle such real life constraints in a clustering algorithm is important.

In this paper, we propose an efficient spatial clustering technique, CPO, which considers the presence of obstacles. The algorithm divides the spatial area into rectangular cells and labels each cell as either *dense* or *non-dense* (according to the number of points in this cell) and also as either *obstructed* (intersects any obstacle) or *non-obstructed*. For each *obstructed* cell, the algorithm finds a number of *non-obstructed* sub-cells. Then it finds the *dense* regions of *non-obstructed* cells or sub-cells by a breadth-first search as the required clusters and determines a center for each region.

The proposed algorithm has several advantages over other work [THH01].
1. It handles outliers or noise.
2. It dose not use any randomized search.
3. Instead of specifying the number of desired clusters beforehand, it finds the natural number of clusters in the area.
4. When the data is updated, we do not need to recompute all information in the cell grid. Instead, incremental update can be done.

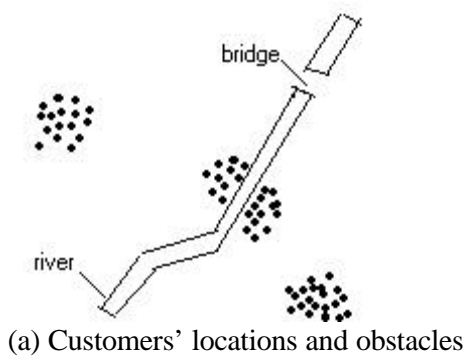
(a) Customers' locations and obstacles

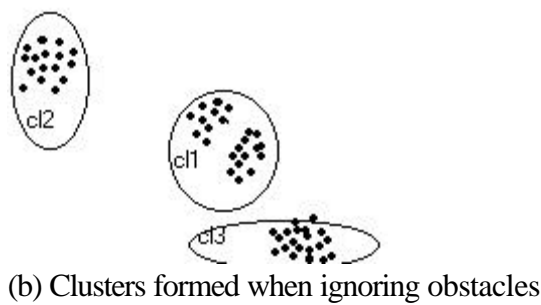
(b) Clusters formed when ignoring obstacles

Figure 1. Planning locations for telephone cabinets

The rest of the paper is organized as follows. We first discuss the related work in Section 2. The proposed algorithm is given in Section 3. In section 4, we analyze the complexity of our algorithm. Section 5 concludes the conclusion.

## 2. RELATED WORK

Many studies [HK00] have been conducted in cluster analysis. These methods can be categorized into partitioning methods [NH94, BFR98], hierarchal methods [ZRL96, GRS98, KHK99], density-based methods [ABKS99, EKSX96, HK98], grid-based methods [WYM97, AGG98, SCA98], and constrained-based methods [THH01]. Methods related to our work are discussed briefly in this section and we emphasize what we believe the limitations, which are addressed by our approach.

### 2.1 COD-CLARANS

COD-CLARANS [THH01] (clustering with obstructed distance based on CLARANS [NH94]) was the first and the only clustering work that consider the presence of obstacle entities in the spatial area. COD-CLARANS takes as input the number, $k$, of the required clusters. That is, the users determine the number of clusters. In some situation, however, determining this number is not easy. COD-CLARANS contains two phases. The first phase breaks the database into several databases and summarizes them individually by grouping the objects in each sub-database in micro-clusters. A micro-cluster is a group of points, which are so close together that they are likely to belong to the same cluster. The second phase is the clustering stage. The algorithm first randomly selects $k$ points as centers for the required $k$ clusters and then tries to find better solutions. A basic difference between CLARANS and COD-CAARANS is that COD-CLARANS computes obstructed distances between the center of a cluster and its objects. The obstructed distance between two points is defined as the length of the shortest Euclidean path between the two points without cutting through any obstacles.

Although COD-CLARANS generates good clustering results, there are several major problems with this algorithm. First, the quality of the results cannot be guaranteed when the number of points, *N*, is large since the randomized search is used in the algorithm to determine initial centers for the clusters and then to refine those centers. Second, COD-CLARANS takes as an input the number of the desired clusters and another integer, which determine the number of maximum tries to refine a center, but both numbers are generally unknowns in realistic applications. Third, COD-CLARANS can't handle outliers. Forth, when data is updated, we need to run the algorithm from scratch.

## 2.2. STING

STING (STatistical INformation Grid), presented in [WYM97], is a statistical information grid-based approach to spatial data mining. A pyramid-like structure is employed, in which the spatial area is divided recursively into rectangular cells down to certain granularity determined by the data distribution and resolution required by applications. Statistical information for each cell is calculated in bottom up manner and is used to answer queries. The results of such queries are in the form of regions that satisfy the conditions specified in the query. So these resulted regions are a clustering of the data according the conditions specified in the query. When processing a query, the hierarchal structure is examined in a top down manner. Cells are marked as either relevant or non-relevant with certain confidence level using standard statistical tests. Only children cells of relevant cells are examined at the next level. The final result is formed as the union of qualified leave level cells.

## 3. ALGORITHM

In this section, we describe our proposed algorithm in detail. We show the main function of the CPO algorithm in Algorithm 3.1. The algorithm first divides the spatial area into rectangle cells of equal size such that the average number of points in each cell is in the range from several dozens to several thousands. Then, the algorithm labels each cell as either *dense* or *non-dense* (according to the number of points in that cell) and also as either *obstructed* (i.e. intersects any obstacle) or *non-obstructed*. Next, each *obstructed* cell is divided into a number of *non-obstructed* sub-cells. Again each of the new sub-cells is labeled as *dense* or *non-dense* (according to the number of point in this sub-cell). Then the algorithm finds the *dense* regions of *non-obstructed* cells or sub-cells by a breadth-first search. The obtained regions are the required clusters. For each cluster the algorithm finds a center. Finally the algorithm outputs the clusters with their centers.

**Algorithm 3.1 CPO.**
**Input:** A set of *N* objects (points) and a set of polygon obstacles in a spatial area *S*.
**Output:** The relatively *dense* regions in *S*, with a center for each region.
**Method:**

1. Let *La* and *Lo* be the dimensions of the spatial area. Determine two numbers, *x*, *y*, such that $x/la = y/lo$ and the average number of points in each cell, *t*, given by $\frac{N}{(la/x)*(lo/y)}$ ranges from several dozens to several thousands.
2. Divide the spatial area into $(la/x)*(lo/y)$ rectangular cells that have equal areas by dividing the longitude and latitude of the spatial area into $(la/x)(=(lo/y))$ equal segments.
3. For each cell, *c*, we determine the following parameters:
   $n_c$ : the number of the objects in the cell.
   $m_c$ : the mean of points in the cell.
4. For each cell, *c*, if $n_c \geq t$, then label *c* as *dense,* otherwise label c as *non-dense*.
5. For each obstacle, *O*, all cells that intersect *O* are labeled as *obstructed*.
6. For each *obstructed* cell, *c*, apply algorithm 3.2 to find *non-obstructed* sub-cells in *c*.

7. For each sub-cell, *sc*, obtained in step 6, if $n_{sc}/t \geq p_{sc}$, then label *sc* as *dense*. Otherwise label *sc* as *non-dense*.
8. For each *dense*, *non-obstructed* cell that is not, previously, processed in the current step, or *dense* sub-cell that also is not processed before in the current step, we examine its neighboring *non-obstructed* cells or sub-cells to see if the average number of points in a cell within this small area is greater than or equal *t*. If so, this area is marked and all *dense* cells or sub-cells that are just examined are put into a queue. Remove from the queue each *dense* cell or sub-cell that has been examined before in a previous iteration. Each time we take one cell or sub-cell from the queue and repeat the same procedure. When the queue is empty, we have identified one region, *cl*.
9. For each region, *cl*, constructed in step 8, apply algorithm 3.3 to find a center for *cl*.
10. Output the constructed regions with their centers.

**Algorithm 3.2** *Find_non-obstruced_sub-cell(c).*
**Input:** an *obstructed* cell, *c*.
**Output:** a number of *non-obstructed* sub-cells in *c*.
**Method:**

1. Divide the cell *c* into a number of small pieces of equal areas in the same way as we divide the spatial area such that the average number of the points in each piece (smaller than the average number of points in a cell inside the spatial area) is in the range from several dozens to several hundreds.
2. For each small piece, *p*, label *p* as either *obstructed* (i.e. intersects any obstacle) or *non-obstructed*.
3. For each *non-obstructed* piece, *p*, that is not marked before in this step, the area constituted from *p* and its *non-obstructed* neighbors is marked and all *non-obstructed* neighbors we just examined are put into a queue. Each time, we take one piece from the queue and repeat the same procedure except that those *non-obstructed* pieces that are not marked before are enqueued. When the queue is empty, we have identified one sub-cell.
4. For each sub-cell, *sc*, obtained in step 3, we determine the following parameters:
   $n_{sc}$: the number of the objects in the sub-cell.
   $m_c$: the mean of points in the sub-cell.
   $P_{sc}$: the percentage of the area covered by *sc* from *c*.
5. Output the sub-cells formed in step 3 with their parameters.

**Algorithm 3.3** *Find_center(cl)*
**Input:** A cluster, *cl*.
**Output:** a center for the cluster *cl*.
**Method:**

1. Calculate the mean, *m*, of all the points in *cl*.
2. If this mean is not in any obstacle, then return *m*.
3. In the case that the mean lies in an obstacle, for each cell, *c*, with mean $m_c$ in the cluster, find cost(*c*) which is $\sum_{i \in cl}(d'(m_c, m_i))^2$, where *i* is a cell in the cluster and $d'(m_c, m_i)$ is the obstructed distance [THH01] between $m_c$ and $m_i$, which defined as the shortest Euclidean path from $m_c$ to $m_i$ that does not go through any obstacle.
4. Return the mean of the cell with the minimum cost.

## 4. ANALYSIS OF THE ALGORITHM

In this section we analysis the complexity of the CPO algorithm. We first need the following definition.

**Definition (Visibility Graph):** Let *B* be the set of obstacles. Each obstacle *O* is represented as a polygon. $V_o$ is the set of vertices that determine the obstacle *O*. The visibility graph of the set of obstacles *B* is an undirected graph with set of nodes *V* and set of edges *E*. *V* is the union of all the sets $V_o$ for all obstacles. There is an edge between two vertices $v_i$ and $v_j$ if they are mutually visible. Two vertices are said to be visible if the straight line that connects the vertices does not intersect an obstacle. The following notation are defined for this discussion:

$N$: the number of data objects in the spatial area S.
$m$: the number of the cells in the spatial area.
$m'$: the number of the cells in the spatial area, if we divide the spatial area such that the number of points in each cell is in the range from several dozens to several hundreds, $m<m'<<N$.

Steps 1 and 2, take a constant time. In step 3, we need to scan points in the spatial area only once so step 3 takes $O(N)$ time. In the worst case, step 4, and step 5 as well, may need to scan all cells, so it takes $O(m)$ time. So the running time for the first five steps is $O(m)+O(m)+O(N)$. Since $m<<N$ the complexity is $O(N)$ time.

### 4.1. The complexity of steps 6, 7 and 8

In the worst case, all cells may be obstructed so step 6 will be applied on all cells. But this is equivalent to re-execute steps 1 and 2 on the spatial area such that the number of points in every cell in the new grid is in the range from several dozens to several hundreds. The number of cells in the new grid is $m'$. Afterwards, to go through step 6, we need to scan all cells in the new grid to label them as *obstructed* or *non-obstructed*, form the *non-obstructed* sub-cells and compute the parameter $p$ for each sub-cell and this scan takes $O(m')$ time and then scan the data points again to determine the other parameters to each sub-cell which takes $O(N)$ time. So, totally step 6 takes $O(N)+O(m')$ time. In step 7, we need to scan all the formed sub-cells to label them as *dense* or *non-dense* but this takes $O(m')$ time. Step 8 in the worst case takes $O(m')$ time.

### 4.2. The complexity of step 9

The worst case running time of step 9 is $O(m|V|^2)+O(m^2|V|)$. The worst case is when the mean of the points in the region is inside an obstacle, we will need to scan all cells in the cluster and find the cell whose mean point is the nearest to the mean points of all the other cells in the region. For each cell, to determine the sum of the obstructed distances between its mean point and the mean points of the other cells in the cluster, this takes $O(|V|^2)+O(m|V|)$. Calculation of this sum includes the following two operations. First, for the mean point of this cell a new Extended Dijkstra's tree is built. The construction of the Extended Dijkstra's tree takes $O(|V|^2)$ time, which is the same as the Dijkstra algorithm[O'R98]. Second, for each of the cells in the regions, we need to look up its visibility information. This takes $O(|V|)$ to each cell, so the second operation takes as a total $O(m|V|)$. Now, since the calculation of summation is needed for each cell in the cluster, the total complexity of this step is $O(m|V|^2)+O(m^2|V|)$.

### 4.3 The total complexity

The total complexity of CPO algorithm is the sum of the running time of all the steps:

$O(CPO) = O(N) +O(N)+O(m')+ O(m')+$
$\qquad O(m')+ O(m|V|^2) + O(m^2|V|)$
$\qquad =O(N) + O(m|V|^2) + O(m^2|V|)$.

### 5. CONCLUSION

In this paper, we introduced a new approach to spatial clustering in the presence of obstacles, which overcomes the disadvantages of the previous work. The new approach works as follows. It first divides the spatial area into rectangle cells of equal size such that the average number of points in each cell is in the range from several dozens to several thousands. Then, the algorithm labels each cell as either *dense* or *non-dense* (according to the number of points in that cell) and also as either *obstructed* (i.e. intersects any obstacle) or *non-obstructed*. Next, each *obstructed* cell is divided into a number of *non-obstructed* sub-cells. Again each of the new sub-cells is labeled as *dense* or *non-dense* (according to the number of point in this sub-cell). Then the algorithm finds the *dense* regions of *non-obstructed* cells or sub-cells by a breadth-first search. The obtained regions are the required clusters. For each cluster the algorithm finds a center. Finally the algorithm outputs the clusters with their centers.

We also gave a complexity analysis of the proposed algorithm.